\documentclass[
superscriptaddress,
 amsmath,amssymb,
 aps,
 pre,
 twocolumn,
]{revtex4}

\usepackage[english]{babel}
\usepackage{graphicx}
\usepackage{dcolumn}
\usepackage{bm}
\usepackage{color}
\usepackage[mathlines]{lineno}
\usepackage[]{amsmath}
\newcommand{\B}[1]{{\bm{#1}}}
\def\<{\langle}
\def\>{\rangle}

\def\Pe{\rm Pe}

\begin{document}
\title{Frictional Active Brownian Particles}
\author{Pin Nie}
\affiliation{ School of Physical and Mathematical Science, Nanyang Technological University, Singapore}
\affiliation{Singapore-MIT Alliance for Research and Technology, Singapore}
\author{Joyjit Chattoraj}
\affiliation{ School of Physical and Mathematical Science, Nanyang Technological University, Singapore}
\author{Antonio Piscitelli}
\affiliation{ School of Physical and Mathematical Science, Nanyang Technological University, Singapore}
\affiliation{CNR--SPIN, Dipartimento di Scienze Fisiche,
Universit\`a di Napoli Federico II, I-80126, Napoli, Italy}
\author{Patrick Doyle}
\affiliation{Singapore-MIT Alliance for Research and Technology, Singapore}
\affiliation{Department of Chemical Engineering, Massachusetts Institute of Technology, Cambridge, Massachusetts,
USA}
\author{Ran Ni}
\email{r.ni@ntu.edu.sg}
\affiliation{School of Chemical and Biomedical Engineering, Nanyang Technological University}
\author{Massimo Pica Ciamarra}
\email{massimo@ntu.edu.sg}
\affiliation{ School of Physical and Mathematical Science, Nanyang Technological University, Singapore}
\affiliation{CNR--SPIN, Dipartimento di Scienze Fisiche,
Universit\`a di Napoli Federico II, I-80126, Napoli, Italy}

\date{\today}

\begin{abstract}
Frictional forces affect the rheology of hard-sphere colloids, at high shear rate.
Here we demonstrate, via numerical simulations, that they also affect the dynamics of active Brownian particles, and their motility induced phase separation. 
Frictional forces increase the angular diffusivity of the particles, in the dilute phase, and prevent colliding particles from resolving their collision by sliding one past to the other.
This leads to qualitatively changes of  motility induced phase diagram in the volume-fraction motility plane. 
While frictionless systems become unstable towards phase separation as the motility increases only if their volume fraction overcomes a threshold, frictional system become unstable regardless of their volume fraction.
These results suggest the possibility of controlling the motility induced phase diagram by tuning the roughness of the particles.
\end{abstract}
\maketitle

\section{Introduction}
The interaction force between macroscopic objects in direct physical contact has a frictional component. 
In colloidal hard-sphere suspensions, direct interparticle contacts are generally suppressed by frictionless repulsive forces, of electrostatic or polymeric origin, which are needed to stabilize the suspension, as well as by lubrication forces~\cite{Israelachvili2011}.
Hence, in these systems, frictional forces are generally negligible. 
Recent results~\cite{Guy2015,Clavaud2017,Hsu2018,Kawasaki2018} have however shown that in colloidal systems under shear, frictional force might become relevant.
This occurs as the relative velocity between contacting particles is of order $\sigma \dot \gamma$, with $\sigma$ particle diameter and $\dot \gamma$ the shear rate. 
At large enough $\dot \gamma$, colliding particles become able to overcome their lubrication interaction, entering into direct physical contact. 
The resulting frictional forces are believed to trigger the discontinuous shear thickening~\cite{Guy2015,Clavaud2017,Hsu2018,Kawasaki2018} phenomenology, an abrupt increase of the shear viscosity with the shear rate.

In systems of self-propelled colloidal particles the relative velocity between colliding particles could also be high. 
Therefore, frictional forces could play a role in these systems by affecting their distinguishing feature, which is a motility induced phase separation (MIPS) from a homogeneous state, to one in which a high-density liquid-like state coexists with a low-density gas-like state.
While the physical origin and the features of this transition have been deeply investigated in the last few years, in both numerical model systems~\cite{Redner2013,Wysocki2014,Fily2012} as well as experimental realizations~\cite{Buttinoni2013,Palacci2013,Theurkauff2012,Ginot2018}, the role of frictional forces causing colliding particles to exert torques on each other has been ignored. 

In this manuscript, we investigate the effect of frictional forces on the motility induced phase separation of active spherical Brownian particles (ABP), a prototypical active matter system. 
In this model, hard-sphere like particles of diameter $\sigma$ are equipped with a polarity ${\B n}$ along which they self-propel with active velocity $v_a$. 
The self-propelling directions change as ${\B n}$ undergo rotational Brownian motion, with rotational diffusion coefficient $D_r$.
Thermal noise also acts on the positional degree of freedom.
The motility induced phase separation of this model is controlled by two variables, the volume fraction, $\phi$, and the Peclet number, $\Pe \equiv v_a/(D_r\sigma)$.
Here we show that friction qualitatively affects the dynamical properties of ABPs, in the homogeneous phase, by enhancing the rotational diffusion while suppressing the translational one. 
Because of this, friction qualitative changes the spinodal line marking the limit of stability of the homogeneous phase in the $\phi$-$\Pe$ plane, at high $\Pe$.
While in the absence of friction~\cite{Cates2015} the low-density spinodal line diverges at a finite volume fraction $\phi_m > 0$, in the presence of friction it diverges at $\phi_m \to 0$.
In this respect, friction makes the motility induced phase diagram of ABPs closer to that observed in most active particle systems, including dumbbells~\cite{Suma2014,Petrelli2018} and schematic models such as run-and-tumble particles~\cite{Cates2015},
active Ornstein–Uhlenbeck~\cite{Klamser2018} and Monte-Carlo models~\cite{Levis2014}, and also closer to gas-liquid transition phase diagram in passive systems. 
Since the frictional interaction between colloidal scale particles can be experimentally tuned~\cite{Hsu2018}, our result indicates that it is possible to experimentally modulate the motility induce phase diagram by optimising the particle roughness.

The paper is organised as follows.
After describing our model in Sec.~\ref{sec:model}, we compare in 
Sec.~\ref{sec:homogeneous} the frictionless and the frictional dynamics, in the homogeneous phase, showing that friction suppresses the translational diffusivity while it enhances the rotational one.
The friction dependence of the motility induced phase diagram is discussed in Sec.~\ref{sec:mips}.
Sec.~\ref{sec:separated} discusses the dynamics in the phase separated region, and highlight how friction promotes the stability of active clusters and hence promotes separation.

\section{Numerical Model\label{sec:model}}
We consider two- and three-dimensional suspensions of active spherical Brownian particles (ABPs) with average diameter $\sigma$ (polydispersity: 2.89\%) and mass $m$, in the overdamped limit. 
The equations of motion for the translational and the rotational velocities are 
\begin{eqnarray}
{\B {v}}_{i} &=& \frac{{\B F}_i}{\gamma}+ \frac{F_a}{\gamma} \hat {\B n}_{i} + \sqrt{2D_t^{0}}\B\eta_i^t \\
{\B \omega}_{i} &=& \frac{ {\B T}_i}{\gamma_r} + \sqrt{2D_r^{0}}\B\eta_i^r.
\end{eqnarray}
Here $D_r^{0}$ and $D_t^{0}= D_r^{0}\sigma^2/3$ are the rotational and the translational diffusion coefficients, $\gamma$ is the viscosity, $\gamma_r = \gamma\frac{\sigma^2}{3}$,
$\eta$ is Gaussian white noise variable with $\<\eta\> =0$ and $\<\eta(t)\eta(t')\> = \delta(t-t')$, $F_a$ the magnitude of the active force acting on the particle and $\hat {\B n}_{i}$ its direction,
$\B F_{i} = \displaystyle\sum \B F_{ij}$ and $\B T_{i} = \frac{\sigma_i}{2} \displaystyle\sum (\B {\hat r}_{ij}\times \B F_{ij})$ are the forces and the torques arising from the interparticle interactions.
In the absence of interaction and noise, particles move with velocity $v_a = F_a/\gamma$, and do not rotate.

We use an interparticle interaction model borrowed from the granular community, to model frictional particles.
The interaction force has a normal and a tangential component, $\B F_{ij} = \B f_{ij}^{n} +  \B f_{ij}^{t}$.
The normal interaction is a purely repulsive Harmonic interaction, $\B{f}^n_{ij}=k_n(\sigma_{ij}-r_{ij})\Theta(\sigma_{ij}-r_{ij})\B{ \hat r}_{ij}$, $\Theta(x)$ is the Heaviside function, $\sigma_{ij}=(1/2)(\sigma_i+ \sigma_j)$, $\B r_{ij}=\B r_i-\B r_j$, and $\B r_i$ is the position of particle $i$.
The tangential force is $\B{f}^t_{ij}=k_t\vec{\xi}_{ij}$,  where $\B{\xi}_{ij}$ is the shear displacement, defined as the integral of the relative velocity of the interacting particle at the contact point throughout the contact, and $k_t=\frac{2}{7}k_n$. 
In addition, the magnitude of tangential force is bounded according to Coulomb's condition: $\left|\B {f}^t_{ij}\right|\leq\mu\left|\B {f}^n_{ij}\right|$.
Working in the overdamped limit, we neglect any viscous dissipation in the interparticle interaction.
In the granular model, we also neglect the presence of rolling friction~\cite{Zahra2012} we expect not to qualitatively affect our results, in analogy with recent findings~\cite{Singh2020} on the role of rolling friction on discontinuous shear thickening.
The value of $k_n$ is chosen to work in the hard-sphere limit, the maximum deformation of a particle being of order $\delta/\sigma \leq 5\times 10^{-4}$.
We simulate systems with $N = 10^4$, unless otherwise states, in the overdamped limit, with integration timestep $2\times{10}^{-8}/D_r^{0}$, using peridic boundary conditions.
We have checked that for the considered value of $N$ finite size effects are negligible away from the critical point, in the range of parameters we consider.
Data are collected after allowing the system to reach a steady-state via simulation lasting at least $2\tau$, where $\tau$ is the time at which the diffusive regime is attained we estimate from the study of the mean square displacement.

\section{Dynamics in the homogeneous phase~\label{sec:homogeneous}}
\subsection{Frictional effect on the interparticle collision~\label{sec:collision}}
\begin{figure}[!!t]
\centering
\includegraphics[width=0.45\textwidth]{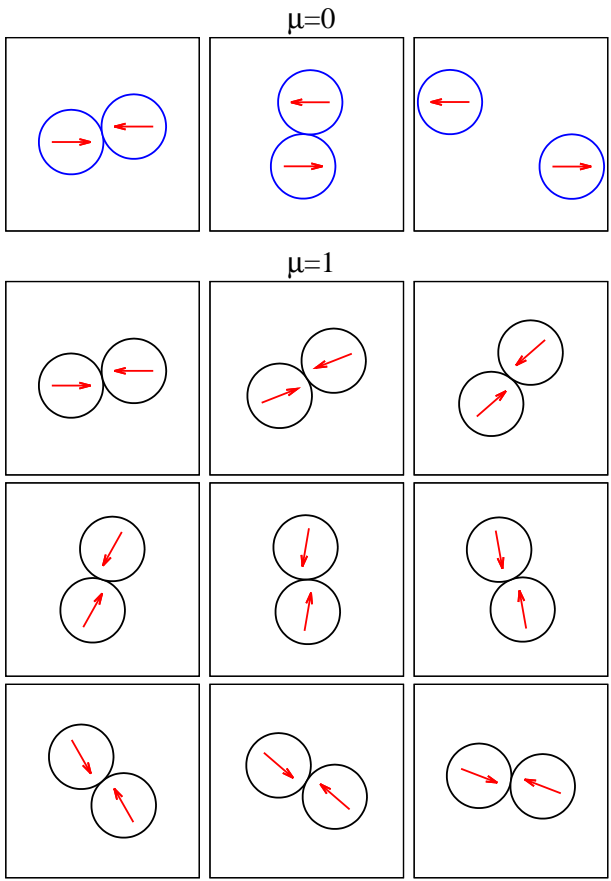}
\caption{
Time evolution of the positions and of the self-propelling directions of two colliding particles, for frictionless (top row) and for frictional (bottom rows) particles. Time evolves from left to right and from top to bottom. In the time interval between consecutive images a free particle move of one diameter. In these simulations, we neglect both translational and rotational noise.
\label{fig:coll}
}
\end{figure}

To appreciate the role of friction on the properties of ABPs, we start considering how friction affects the collision between two particles.
The interparticle force acting between two colliding particles generally has a component parallel to the line joining the centers of the two particles, and a tangential component. 
In the absence of friction, this tangential component allows the particles to slide one past the other to resolve their collision, as illustrated in the upper row of Fig.~\ref{fig:coll}.

In the presence of friction, particles are not free to tangentially slide one past the other. 
Specifically, the tangential shearing induced an opposing frictional force that slows down the motion of the particles, and it induces their rotation. 
In the absence of thermal forces, or equivalently in the $\Pe \to \infty$ limit, the frictional forces cause the particles to rigidly rotate around their contact point, so that they never resolving their collision, as is illustrated in the bottom rows of Fig.~\ref{fig:coll}.
At any finite $\Pe$, the stochastic forces acting on the particles will be able to break their contact, and hence the frictional forces, allowing the particles to resolve their collision.
We do expect, therefore, that friction may affect the physics of ABP at high $\Pe$, inducing a non-negligible rotation of the self-propelling directions of colliding particles. 

\subsection{Dilute phase\label{sec:dilute}}
\begin{figure}[!!t]
\centering
\includegraphics[width=0.52\textwidth]{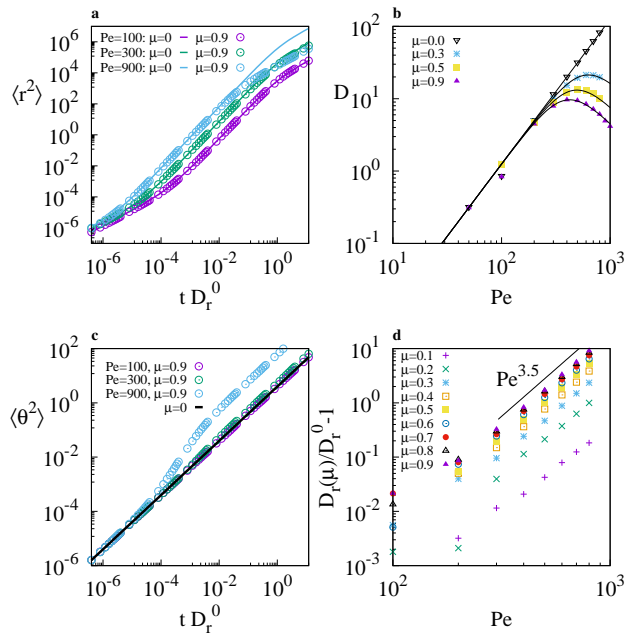}
\caption{
Mean square displacement ({\bf a}) for the frictionless ($\mu = 0$, full lines) and the frictional dynamics ($\mu = 0.9$, symbols) for selected values of the Peclet number.
At large $\Pe$ friction reduces the time at which the dynamics enter the asymptotic diffusive regime, and hence suppresses the diffusion coefficient ({\bf b}).
The mean square angular displacement ({\bf c}) is enhanced by the frictional force, at times larger than the interparticle collision time. This leads to an increase of the rotational diffusivity ({\bf d}) scaling as $\mu^2\Pe^x$, with $x \simeq 3.5$.
Lines in ({\bf b}) are one parameter fits to the theoretical prediction
of Eq.~\ref{eq:Dpred} where $x = 3.5$ is held constant.
In all panels, $\phi = 0.1$.
}
\label{fig:dynamics}
\end{figure}
We start describing the effect of friction on the MIPS, comparing the dynamics of frictionless and frictional systems in the homogeneous phase.
Fig.~\ref{fig:dynamics}{\bf a} illustrates the frictionless and the frictional mean square displacement, at different Peclet numbers, for volume fractions in the gas phase.
In the absence of friction (full lines) the mean square displacement exhibits a crossovers from a diffusive to a super-diffusive regime at $t \simeq 6D_t^{0}/v^2_{a}$, and from the super-diffusive to the asymptotic diffusive regime at $t = 1/D_r^{0}$~\cite{Redner2013}.
In the presence of friction (symbols), similar behaviour is observed, but the system enters the diffusive regime on a smaller time scale.
Consequently, the translational diffusivity is also reduced as illustrated in Fig.~\ref{fig:dynamics}{\bf b}.
This finding is rationalised investigating the mean square angular displacement (\ref{fig:dynamics}{\bf c}) and the dependence of the rotational diffusivity $D_r$ on $\Pe$ (\ref{fig:dynamics}{\bf d}).
Indeed, these quantities clarify that friction enhances the rotational diffusion of the particles, hence reducing the timescale at which the system enters the asymptotic translational diffusive regime.

We rationalize how friction leads to an increase of the rotational diffusivity, considering that in a collision a frictional particle experiences a torque, which induces the rotation of its self-propelling direction.
More quantitatively, in the overdamped limit, the rotation $\Delta\theta_i$ induced by a collision is proportional to the induced torque, and the duration of the contact. 
If the contacts are at their critical Coulomb value, the typical torque magnitude is $\sigma f^t \propto \mu f^n \propto \mu \Pe$, and the mean squared angular displacement induced by a collision  of duration $t_{\rm coll}$ is 
$\langle \Delta\theta_i^2\rangle \propto \mu^2 \Pe^2 t_{\rm coll}^2$.
At low density consecutive torques experienced by a particle are uncorrelated and the number of collisions per unit time is proportional to $\Pe$. Hence, assuming $t_{\rm coll} \propto \Pe^q$, we predict for the rotational diffusivity
\begin{equation}
D_r(\Pe,\mu) = D_r^{0}+ \alpha \mu^2 \Pe^x
\label{eq:Drpred}
\end{equation}
with $x = 3+2q$. 
Our numerical results of Fig.\ref{fig:dynamics}{\bf d} indicate $x \simeq 3.5$. 
These results indicate that the average duration of an interparticle collision slightly grows with the Peclet number, $t_{\rm coll} \propto \Pe^{1/4}$. 
We qualitatively rationalize the dependence of the collision duration on the Peclet number considering that $\Pe$ controls the ratio between the frictional forces, which protract the duration of contacts, and the thermal ones, which eventually allow particles to resolve their collision. 
We have indeed observed in Fig.~\ref{fig:coll} that in the $\Pe \to \infty$ limit collisions are not resolved, so that $t_{\rm coll} = \infty$.

The dependence of the rotational diffusion coefficient on $\Pe$ and on $\mu$ allows also to rationalise the non monotonic behaviour of the diffusivity observed in Fig.~\ref{fig:dynamics}{\bf b}.
Indeed, in the $\phi \to 0$ limit the long time mean square displacement of an active particle is $\Delta r^2(t) = 6D_t^{0}t + \frac{v_a^2}{D_r} t$.
At a small but finite density $\phi$, we therefore expect
\begin{equation}
\label{eq:Dpred}
D(\Pe,\mu) = c(\phi) \left[ D_t^{0}+ \frac{\sigma^2}{6} \frac{(D_r^{0})^2}{D_r(\Pe,\mu)} \Pe^2 \right]
\end{equation}
with $c(\phi)$ a constant of order one, $D_r(\Pe,\mu)$ is given by Eq.~\ref{eq:Drpred}.
Eq.~\ref{eq:Dpred} well describes the data of Fig.~\ref{fig:dynamics}{\bf b}, with $c(\phi=0.1) \simeq 0.8$. Hence, the diffusivity grows as $\Pe^2$ at small $\Pe$, and decreases as $\Pe^{2-x}$ at large $\Pe$.

\section{Frictional MIPS\label{sec:mips}}
\begin{figure}[t!]
\centering
\includegraphics[width=0.5\textwidth]{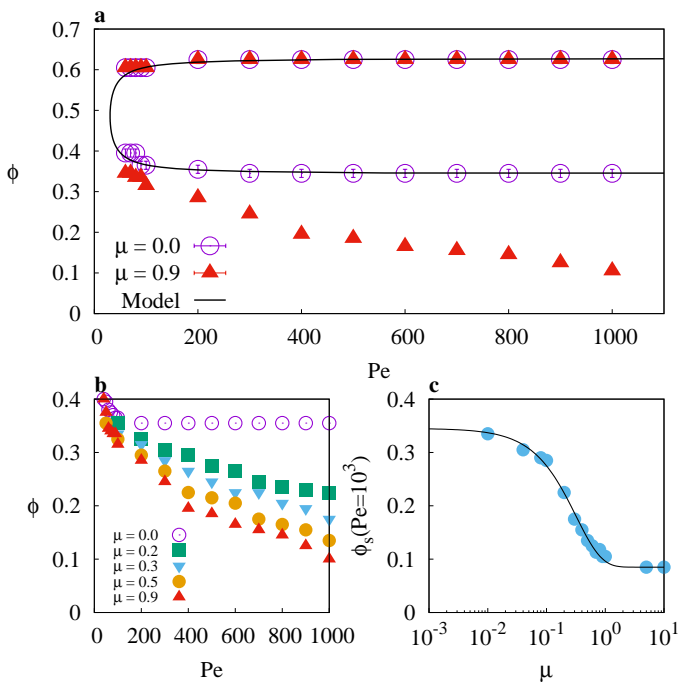}
\caption{
The phase diagrams of frictionless (circles) and of frictional (triangles) ABPs are compared in {(\bf a}). The full line, a parabola, is the theoretical prediction of Ref.~\cite{Nie2020}, for frictionless systems.
At high $\Pe$ the low density critical line saturates to a constant value in the absence of friction,
while it vanishes in the frictional case. This occurs for all values of the friction coefficient, as
shown in ({\bf b}). At a given $\Pe$, the critical volume fraction at which phase separation occurs varies
with friction as $\phi_s(\mu) = \phi_s(\infty) + \Delta\phi_s e^{-\mu/\mu_c}$. For $\Pe = 10^3$
({\bf c}) we find $\phi_s(\infty) = 0.085(3)$, $\Delta\phi_s=0.26(2)$ and $\mu_c=0.32(1)$.
}
\label{fig:friction}
\end{figure}
We have investigated the motility phase diagram as a function of the friction coefficient $\mu$, of the volume fraction $\phi$, and of the 
Peclet number ${\Pe \equiv} v_a\tau_B/\sigma$,
where $v_a$ is the particle velocity in the $\phi \to 0$ limit,
$\sigma$ the average particle diameter and $\tau_B = 1/D_r^{0}$ is the Brownian time,
$D_r^{0}$ being the rotational diffusion coefficient of the self-propelling directions in the absence of friction.
We have determined the phase diagram considering the systems to be phase-separated when the distribution of the local density exhibits two peaks, at the end of relative short simulations.
Indeed, this assures that phase separation has occurred via spinodal decomposition, rather than via nucleation, as we previously verified~\cite{Nie2020}.
We discuss, here, results obtained in three spatial dimensions.

In the absence of friction, our results are in qualitative agreement with previous investigations.
The increase of the Peclet number drives the phase separation of the system, but only for volume fractions above a critical value~\cite{Wysocki2014,Stenhammar2014,Levis2017,Digregorio2018}, as illustrated in Fig.~\ref{fig:friction}{\bf a} (circles).

We highlight how friction influences this scenario by also illustrating in Fig.~\ref{fig:friction}{\bf a} the spinodal line for $\mu = 0.9$ (triangles).
The figure reveals that friction does not appreciably influence the high-density spinodal line, while it strongly affects the low-density line does.
In particular, while $\phi_s(\Pe,\mu = 0)$ reaches a plateau as $\Pe$ increase, $\phi_s(\Pe,0)$ monotonically decreases with $\Pe$.  
A similar effect of friction has been reported on the volume fraction of static granular packing~\cite{Silbert2010,Ciamarra2011}.
Hence, the effect of friction becomes more relevant on increasing $\Pe$, as we anticipated in Sec.~\ref{sec:collision}.

Fig.~\ref{fig:friction}{\bf b} further investigates this dependence illustrating the low-density spinodal line for different values of the friction coefficient. 
Regardless of the $\mu$ value, the spinodal line decreases on increasing $\Pe$ or $\mu$.
Fig.~\ref{fig:friction}{\bf c} illustrates the value of the lower-spinodal line, at $\Pe = 10^3$, as a function of $\mu$.
The figure reveals that this value exponentially decreases with $\mu$, approaching a limiting value.
This exponential dependence is rationalized considering that the frictional forces, whose magnitude scale as $\mu v_a \propto \mu \Pe$, can be disrupted by thermal forces, that have a constant magnitude, through an activated process. 
Since the associated Boltzmann factor $\exp\left(-\mu\Pe\right)$ vanishes in the $\Pe \to \infty$ limit, so thus the spinodal line, for $\mu > 0$.
Hence, the frictionless case appears as a singular one.

\begin{figure}[!!t]
\centering
\includegraphics[width=0.48\textwidth]{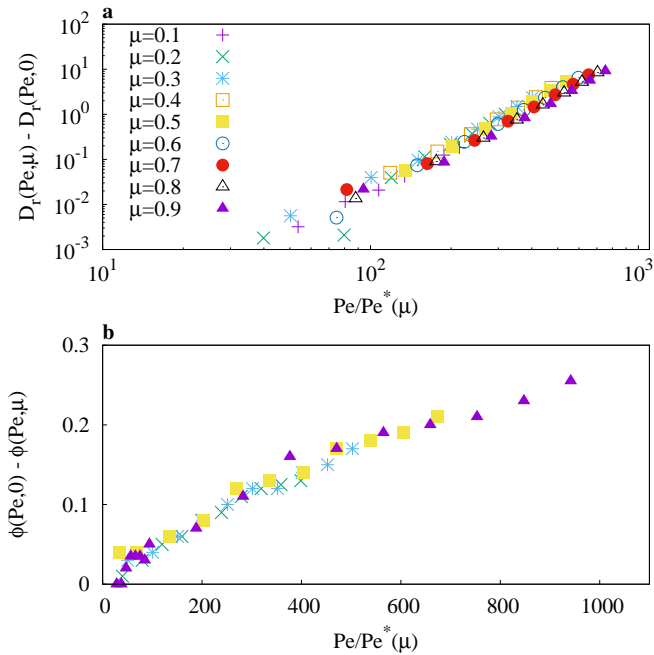}
\caption{Friction leads to an increase of the rotational diffusivity scaling as $D_r(\Pe,\mu) -D_r(Pe,0) \propto \Pe^*\propto \mu^{-4/7}$ ({\bf a}).
The difference between the frictionless and the frictional low density critical lines (see Fig.\ref{fig:friction}{\bf b}) is also controlled by $\Pe^*$ ({\bf b}).}
\label{fig:scalefriction}
\end{figure}

To quantitatively rationalize the interplay between friction and Peclet number, we consider that according to Eq.~\ref{eq:Drpred} frictional forces play a role for $\Pe > \Pe^*\propto \mu^{-4/7}$, as found imposing $\mu (\Pe^*)^x\propto D_r^0$. 
Indeed, we show in Fig.~\ref{fig:scalefriction}{\bf a} that, when plotted vs $\Pe/\Pe^*$, rotational diffusivity data corresponding to different values of the friction coefficient nicely collapse.
Accordingly, the frictional critical line $\phi_c(\Pe,\mu)$ coincides with the frictionless one for $\Pe < \Pe^*(\mu)$, while conversely it deviates from it.
We confirm this expectation in Fig.~\ref{fig:scalefriction}{\bf b},
which illustrates that the distance between the frictionless and the frictional critical lines, $\phi(\Pe,0)-\phi_c(\Pe,\mu)$ scales as $\Pe/\Pe^*(\mu)$.

\section{Dynamics in the phase separated phase~\label{sec:separated}}
\begin{figure*}[!!t]
\centering
\includegraphics[width=1\textwidth]{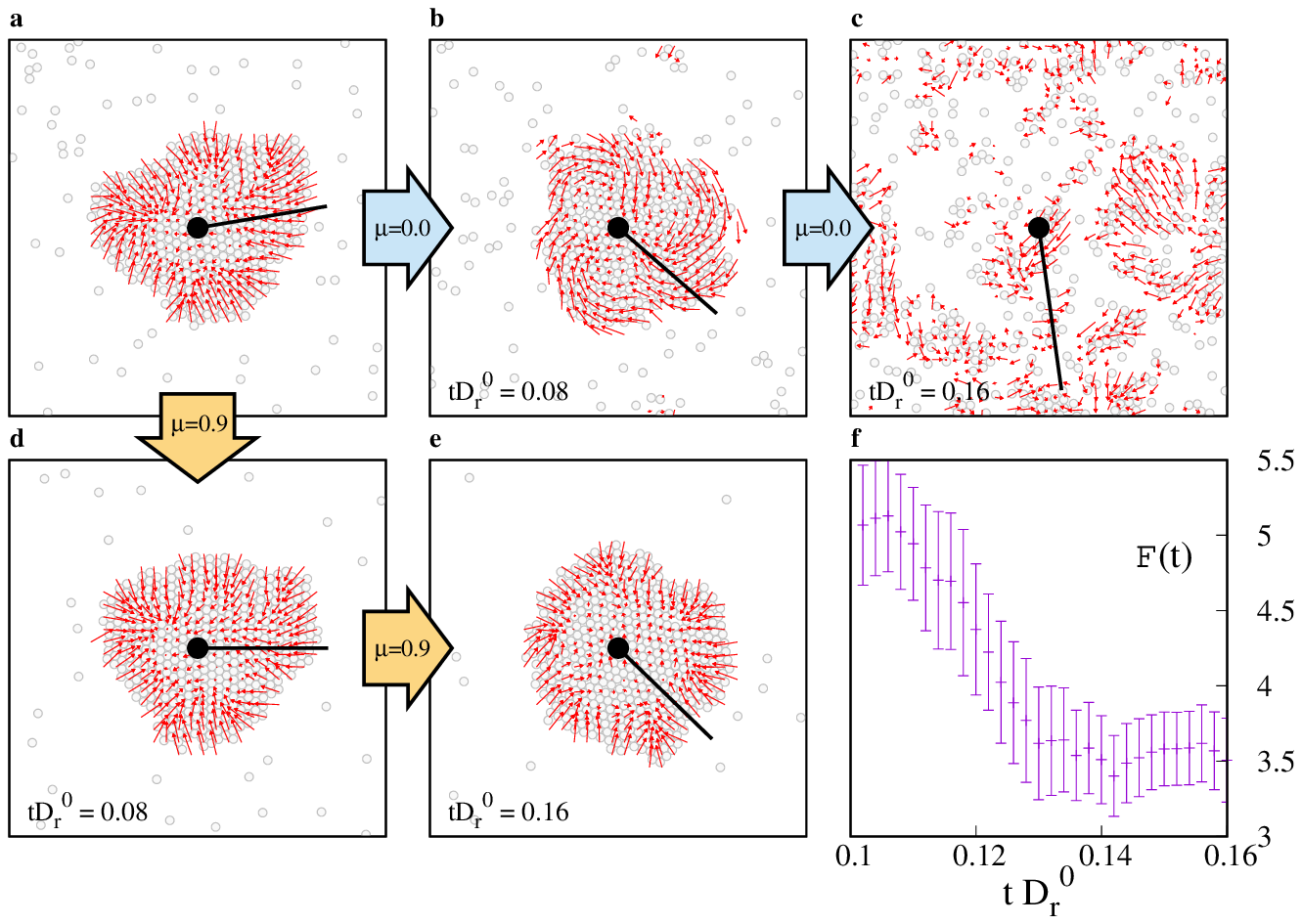}
\caption{
Evolution of a two dimensional cluster of ABPs in the absence ({\bf a,b,c}) and in the presence ({\bf a,d,e}) of frictional forces. The red arrows show the active force field (see methods).
In all plots, the central black circle identifies the position of the particle closer to the centre of mass of the cluster, in the initial configuration.
We emphasize the rotational motion of the cluster, drawing a line connecting the central particle and another particle of the cluster.
Both with and without friction the cluster rigidly rotates around its centre of mass.
In the absence of friction, the rotation of the cluster makes the active velocities parallel to the cluster surface ({\bf b}) inducing a cluster instability ({\bf c}).  
Indeed, we see in panel {\bf f} that, in the absence of friction, as the cluster rotates, the average interaction force that contrasts the motion along the direction of the self-propelling forces decreases, fluctuating around a value characteristic of the homogeneous phase once the cluster breaks.
In the presence of friction, the cluster rotation induces that of the self-propelling directions, and the cluster remains stable ({\bf d,e}).
For these illustrative two-dimensional simulations $N = 500$, $\Pe = 500$, $\phi = 0.2$ and $\mu = 0.0; 0.9$.
\label{fig:rotation}}  
\end{figure*}  

Friction significantly stabilizes phase-separated configurations, thus expanding the coexistence region in the $\phi$-$\Pe$ plane.
To understand how friction stabilized an active cluster, we start considering a frictional simulation for parameter values at which phase separation occurs, $\Pe = 500$, $\phi = 0.2$ and $\mu = 0.9$.
Besides, for ease of visualization, we consider a small two-dimensional system, with $N = 500$ particles, so that in the steady state we readily observe the formation of a single cluster.
This is illustrated in Fig.~\ref{fig:rotation}a.
In this and the other panels of Fig.~\ref{fig:rotation}, we also illustrate the active velocity field evaluated on a square grid with lattice spacing $\sim 1.5 \sigma$.
We associate to each grid point the average active velocity of the particles in a circle of radius $\simeq 2.2 \sigma$. In the figure, we show the values of the field on the grid points that average over at least $5$ particles.
We use the configuration Fig.~\ref{fig:rotation}a as initial configuration of two different simulations, a frictionless ($\mu = 0$) and a frictional one ($\mu = 0.9)$.

Fig.~\ref{fig:rotation}{\bf a}-{\bf c} show that in the absence of friction the cluster becomes unstable and disintegrates upon rotation.
The instability occurs as the rotation makes the self-propelling velocities tangential to the cluster surface.
We have verified that the system becomes macroscopically unstable by investigating the magnitude of the forces which oppose the motion of the particles in their self-propelling directions
$\mathcal F(t) = - \frac{1}{NF_a}\frac{d}{d\alpha} \left. U({\B r}(t) + \alpha {\B {\hat n}}) \right|_0$, where $U$ is the elastic energy of the system and $\B {\hat n}$ is the director of the active velocity field, normalized by the magnitude $F_a$ of the active force acting on each particle and by the number of particles $N$.
Fig.~\ref{fig:rotation}{\bf f} shows that $\mathcal F(t)$ quickly decreases as the cluster rotates, reflecting the development of the instability.

Friction promotes phase separation by suppressing this rotational induced instability of active clusters. This is illustrated in Fig.~\ref{fig:rotation}{\bf a},{\bf d}-{\bf e}.
The frictional cluster is stable because its rotation induces that of the self-propelling directions of its particles. 
Indeed, we observe in Fig.~\ref{fig:rotation}{\bf a},{\bf d}-{\bf e} the active velocities to always point towards the centre of the cluster. 
This tendency is more pronounced the higher the frictional forces, and hence at large $\mu$ and large $\Pe$.
As a consequence of this process, while in the absence of friction active clusters quickly disintegrate as they start rotating, in frictional ABPs one observe long-lasting active clusters, that performs many revolutions before eventually breaking apart. 
In this respect, frictional ABPs behave as active 
dumbells~\cite{Suma2014,Petrelli2018}.

From the decay of $\mathcal{F}(t)$ it is possible to extract the lifetime of the considered cluster. 
For the cluster illustrated in Fig.~\ref{fig:rotation}, This lifetime is $\simeq 0.12 D_r^0$ in the absence friction, as apparent from panel {\bf f}.
For this cluster, we have observed the lifetime to quickly grow as the friction coefficient increases, reaching values beyond our simulation capabilities for $\mu \simeq 0.2$. 
This result strongly suggests that frictional clusters breaks as thermal fluctuations succeed inducing the relative rotation of contacting cluster.

\section{Discussion}
We have recently investigated the stability phase diagram of frictionless active Brownian particles, and rationalized its spinodal line within a kinetic model.
In this model, the finite value of the lower-spinodal line in the $\Pe$ limit, in the absence of friction, results from the competition of two processes.
Phase separation is promoted by the collisions of the particles in the homogeneous phase, which may trigger the agglomeration. 
Being related to the particle velocity, this process leads to a flux of particles $j_{g} \propto \Pe$ from the dilute to the dense phase.
The dilute phase is promoted by the process that allow particles resolving their collisions. At low $\Pe$, particles mainly resolve their collision by rotating their self-propelling direction. In the high-$\Pe$ limit we are interested, conversely, particles resolve their collisions by sliding one past to the other, as we illustrated in Fig.~\ref{fig:coll}, before their self-propelling direction changes.
This sliding-detaching mechanisms, also driven by the motility,  leads to a flux of particles from the dense to the less-dense phase, $j_{\rm sd} \propto \Pe$. 
Since both $j_{g}$ and $j_{\rm sd}$ are proportional to the Peclet number, the spinodal line results not to depend on it.

Within this context, the role of friction is rationalized considering its influence on these contrasting fluxes. 
In Sec.~\ref{sec:dilute}, we have demonstrated that friction influences the homogeneous phase by increasing the rotational diffusivity. 
This increase does not affect the typical velocity of the particles, and hence the  flux $j_{g}$. 
On the other hand, we have found friction to suppressed the ability of two particles to slide one past to the other, to resolve their collision, in Sec.~\ref{sec:collision}.
Similarly, in the phase separate phase, friction stabilizes active cluster, which would conversely disintegrate upon rotation, as discussed in Sec.~\ref{sec:separated}. 
The balance between $j_{g}$ and $j_{sd}$, therefore, leads to a friction depend spinodal line with the coexistence region widening on increasing the friction coefficient.

That the suppression of the sliding detaching mechanism leads to a widening of the coexistence region is consistent with previous finding. 
In the context of frictionless spherical particles, the sliding detaching mechanisms is suppressed when the dynamics is investigated via Monte Carlo simulations, which are unable to account for the cooperative displacement of colliding particles. Consistently, in these simulations the spinodal line is found to vanish in the $\Pe \to \infty$ limit~\cite{Levis2014,Klamser2018}.
The sliding detaching mechanisms is also suppressed in system of active anisotropic particles. 
Indeed, these particles cannot rotate independently when in a dense cluster, which implies that an active cluster of anisotropic particles does not destabilize when rotating, as in frictional spherical particles (see Fig.~\ref{fig:rotation}). Consistently, the spinodal line of frictionless dumbells does also vanish~\cite{Suma2014,Petrelli2018} in the $\Pe \to \infty$ limit. 

Interestingly, we notice that long-lived rotating clusters have also been observed in experiments of active thermophoretic particles~\cite{Theurkauff2012,Palacci2013,Buttinoni2013}.
These clusters might perform several revolutions before restructuring.
While it is understood that these clusters might be stabilised by the
attractive phoretic attraction between the particles~\cite{Theurkauff2012,Palacci2013,Liebchen2019}, it has been suggested that this attraction is not always present~\cite{Buttinoni2013}.
In these circumstances, friction might be a concurring stabilising factor, as one could experimentally ascertain investigating whether the self-propelling directions of the particles rotate with the cluster itself.

Regardless, frictional forces could be enhanced by acting on the roughness of the particles~\cite{Hsu2018}, suggesting that friction could be used as a control parameter to experimentally tune the motility induced phase diagram.

We acknowledge support from the Singapore Ministry of Education through the Academic Research Fund MOE2017-T2-1-066 (S), and are grateful to the National Supercomputing Centre (NSCC) for providing computational resources.


\begin{thebibliography}{26}%
\makeatletter
\providecommand \@ifxundefined [1]{%
 \@ifx{#1\undefined}
}%
\providecommand \@ifnum [1]{%
 \ifnum #1\expandafter \@firstoftwo
 \else \expandafter \@secondoftwo
 \fi
}%
\providecommand \@ifx [1]{%
 \ifx #1\expandafter \@firstoftwo
 \else \expandafter \@secondoftwo
 \fi
}%
\providecommand \natexlab [1]{#1}%
\providecommand \enquote  [1]{``#1''}%
\providecommand \bibnamefont  [1]{#1}%
\providecommand \bibfnamefont [1]{#1}%
\providecommand \citenamefont [1]{#1}%
\providecommand \href@noop [0]{\@secondoftwo}%
\providecommand \href [0]{\begingroup \@sanitize@url \@href}%
\providecommand \@href[1]{\@@startlink{#1}\@@href}%
\providecommand \@@href[1]{\endgroup#1\@@endlink}%
\providecommand \@sanitize@url [0]{\catcode `\\12\catcode `\$12\catcode
  `\&12\catcode `\#12\catcode `\^12\catcode `\_12\catcode `\%12\relax}%
\providecommand \@@startlink[1]{}%
\providecommand \@@endlink[0]{}%
\providecommand \url  [0]{\begingroup\@sanitize@url \@url }%
\providecommand \@url [1]{\endgroup\@href {#1}{\urlprefix }}%
\providecommand \urlprefix  [0]{URL }%
\providecommand \Eprint [0]{\href }%
\providecommand \doibase [0]{https://doi.org/}%
\providecommand \selectlanguage [0]{\@gobble}%
\providecommand \bibinfo  [0]{\@secondoftwo}%
\providecommand \bibfield  [0]{\@secondoftwo}%
\providecommand \translation [1]{[#1]}%
\providecommand \BibitemOpen [0]{}%
\providecommand \bibitemStop [0]{}%
\providecommand \bibitemNoStop [0]{.\EOS\space}%
\providecommand \EOS [0]{\spacefactor3000\relax}%
\providecommand \BibitemShut  [1]{\csname bibitem#1\endcsname}%
\let\auto@bib@innerbib\@empty
\bibitem [{\citenamefont {Israelachvili}(2011)}]{Israelachvili2011}%
  \BibitemOpen
  \bibfield  {author} {\bibinfo {author} {\bibfnamefont {J.~N.}\ \bibnamefont
  {Israelachvili}},\ }\href@noop {} {{\selectlanguage {English}\emph {\bibinfo
  {title} {{Intermolecular and Surface Forces, Third Edition: Revised Third
  Edition}}}}},\ \bibinfo {edition} {3rd}\ ed.\ (\bibinfo  {publisher}
  {Academic Press},\ \bibinfo {address} {Burlington, MA},\ \bibinfo {year}
  {2011})\ p.\ \bibinfo {pages} {704}\BibitemShut {NoStop}%
\bibitem [{\citenamefont {Guy}\ \emph {et~al.}(2015)\citenamefont {Guy},
  \citenamefont {Hermes},\ and\ \citenamefont {Poon}}]{Guy2015}%
  \BibitemOpen
  \bibfield  {author} {\bibinfo {author} {\bibfnamefont {B.~M.}\ \bibnamefont
  {Guy}}, \bibinfo {author} {\bibfnamefont {M.}~\bibnamefont {Hermes}},\ and\
  \bibinfo {author} {\bibfnamefont {W.~C.~K.}\ \bibnamefont {Poon}},\
  }\bibfield  {title} {\bibinfo {title} {{Towards a Unified Description of the
  Rheology of Hard-Particle Suspensions}},\ }\href@noop {} {\bibfield
  {journal} {\bibinfo  {journal} {Phys Rev Lett}\ }\textbf {\bibinfo {volume}
  {115}},\ \bibinfo {pages} {088304} (\bibinfo {year} {2015})}\BibitemShut
  {NoStop}%
\bibitem [{\citenamefont {Clavaud}\ \emph {et~al.}(2017)\citenamefont
  {Clavaud}, \citenamefont {B{\'{e}}rut}, \citenamefont {Metzger},\ and\
  \citenamefont {Forterre}}]{Clavaud2017}%
  \BibitemOpen
  \bibfield  {author} {\bibinfo {author} {\bibfnamefont {C.}~\bibnamefont
  {Clavaud}}, \bibinfo {author} {\bibfnamefont {A.}~\bibnamefont
  {B{\'{e}}rut}}, \bibinfo {author} {\bibfnamefont {B.}~\bibnamefont
  {Metzger}},\ and\ \bibinfo {author} {\bibfnamefont {Y.}~\bibnamefont
  {Forterre}},\ }\bibfield  {title} {\bibinfo {title} {{Revealing the
  frictional transition in shear-thickening suspensions.}},\ }\href@noop {}
  {\bibfield  {journal} {\bibinfo  {journal} {Proceedings of the National
  Academy of Sciences of the United States of America}\ }\textbf {\bibinfo
  {volume} {114}},\ \bibinfo {pages} {5147} (\bibinfo {year}
  {2017})}\BibitemShut {NoStop}%
\bibitem [{\citenamefont {Hsu}\ \emph {et~al.}(2018)\citenamefont {Hsu},
  \citenamefont {Ramakrishna}, \citenamefont {Zanini}, \citenamefont
  {Spencer},\ and\ \citenamefont {Isa}}]{Hsu2018}%
  \BibitemOpen
  \bibfield  {author} {\bibinfo {author} {\bibfnamefont {C.-P.}\ \bibnamefont
  {Hsu}}, \bibinfo {author} {\bibfnamefont {S.~N.}\ \bibnamefont
  {Ramakrishna}}, \bibinfo {author} {\bibfnamefont {M.}~\bibnamefont {Zanini}},
  \bibinfo {author} {\bibfnamefont {N.~D.}\ \bibnamefont {Spencer}},\ and\
  \bibinfo {author} {\bibfnamefont {L.}~\bibnamefont {Isa}},\ }\bibfield
  {title} {\bibinfo {title} {{Roughness-dependent tribology effects on
  discontinuous shear thickening.}},\ }\href@noop {} {\bibfield  {journal}
  {\bibinfo  {journal} {Proceedings of the National Academy of Sciences of the
  United States of America}\ }\textbf {\bibinfo {volume} {115}},\ \bibinfo
  {pages} {5117} (\bibinfo {year} {2018})}\BibitemShut {NoStop}%
\bibitem [{\citenamefont {Kawasaki}\ and\ \citenamefont
  {Berthier}(2018)}]{Kawasaki2018}%
  \BibitemOpen
  \bibfield  {author} {\bibinfo {author} {\bibfnamefont {T.}~\bibnamefont
  {Kawasaki}}\ and\ \bibinfo {author} {\bibfnamefont {L.}~\bibnamefont
  {Berthier}},\ }\bibfield  {title} {\bibinfo {title} {{Discontinuous shear
  thickening in Brownian suspensions}},\ }\href@noop {} {\bibfield  {journal}
  {\bibinfo  {journal} {Physical Review E}\ }\textbf {\bibinfo {volume} {98}},\
  \bibinfo {pages} {012609} (\bibinfo {year} {2018})}\BibitemShut {NoStop}%
\bibitem [{\citenamefont {Redner}\ \emph {et~al.}(2013)\citenamefont {Redner},
  \citenamefont {Hagan}, \citenamefont {Baskaran},\ and\ \citenamefont
  {Fisher}}]{Redner2013}%
  \BibitemOpen
  \bibfield  {author} {\bibinfo {author} {\bibfnamefont {G.~S.}\ \bibnamefont
  {Redner}}, \bibinfo {author} {\bibfnamefont {M.~F.}\ \bibnamefont {Hagan}},
  \bibinfo {author} {\bibfnamefont {A.}~\bibnamefont {Baskaran}},\ and\
  \bibinfo {author} {\bibfnamefont {M.}~\bibnamefont {Fisher}},\ }\bibfield
  {title} {\bibinfo {title} {{Structure and Dynamics of a Phase-Separating
  Active Colloidal Fluid}},\ }\href@noop {} {\bibfield  {journal} {\bibinfo
  {journal} {Phys Rev Lett}\ }\textbf {\bibinfo {volume} {110}},\ \bibinfo
  {pages} {055701} (\bibinfo {year} {2013})}\BibitemShut {NoStop}%
\bibitem [{\citenamefont {Wysocki}\ \emph {et~al.}(2014)\citenamefont
  {Wysocki}, \citenamefont {Winkler},\ and\ \citenamefont
  {Gompper}}]{Wysocki2014}%
  \BibitemOpen
  \bibfield  {author} {\bibinfo {author} {\bibfnamefont {A.}~\bibnamefont
  {Wysocki}}, \bibinfo {author} {\bibfnamefont {R.~G.}\ \bibnamefont
  {Winkler}},\ and\ \bibinfo {author} {\bibfnamefont {G.}~\bibnamefont
  {Gompper}},\ }\bibfield  {title} {\bibinfo {title} {{Cooperative motion of
  active Brownian spheres in three-dimensional dense suspensions}},\
  }\href@noop {} {\bibfield  {journal} {\bibinfo  {journal} {EPL (Europhysics
  Letters)}\ }\textbf {\bibinfo {volume} {105}},\ \bibinfo {pages} {48004}
  (\bibinfo {year} {2014})}\BibitemShut {NoStop}%
\bibitem [{\citenamefont {Fily}\ and\ \citenamefont
  {Marchetti}(2012)}]{Fily2012}%
  \BibitemOpen
  \bibfield  {author} {\bibinfo {author} {\bibfnamefont {Y.}~\bibnamefont
  {Fily}}\ and\ \bibinfo {author} {\bibfnamefont {M.~C.}\ \bibnamefont
  {Marchetti}},\ }\bibfield  {title} {\bibinfo {title} {{Athermal phase
  separation of self-propelled particles with no alignment}},\ }\href@noop {}
  {\bibfield  {journal} {\bibinfo  {journal} {Physical Review Letters}\
  }\textbf {\bibinfo {volume} {108}},\ \bibinfo {pages} {235702} (\bibinfo
  {year} {2012})}\BibitemShut {NoStop}%
\bibitem [{\citenamefont {Buttinoni}\ \emph {et~al.}(2013)\citenamefont
  {Buttinoni}, \citenamefont {Bialk{\'{e}}}, \citenamefont {K{\"{u}}mmel},
  \citenamefont {L{\"{o}}wen}, \citenamefont {Bechinger},\ and\ \citenamefont
  {Speck}}]{Buttinoni2013}%
  \BibitemOpen
  \bibfield  {author} {\bibinfo {author} {\bibfnamefont {I.}~\bibnamefont
  {Buttinoni}}, \bibinfo {author} {\bibfnamefont {J.}~\bibnamefont
  {Bialk{\'{e}}}}, \bibinfo {author} {\bibfnamefont {F.}~\bibnamefont
  {K{\"{u}}mmel}}, \bibinfo {author} {\bibfnamefont {H.}~\bibnamefont
  {L{\"{o}}wen}}, \bibinfo {author} {\bibfnamefont {C.}~\bibnamefont
  {Bechinger}},\ and\ \bibinfo {author} {\bibfnamefont {T.}~\bibnamefont
  {Speck}},\ }\bibfield  {title} {\bibinfo {title} {{Dynamical Clustering and
  Phase Separation in Suspensions of Self-Propelled Colloidal Particles}},\
  }\href@noop {} {\bibfield  {journal} {\bibinfo  {journal} {Phys. Rev. Lett.}\
  ,\ \bibinfo {pages} {238301}} (\bibinfo {year} {2013})}\BibitemShut {NoStop}%
\bibitem [{\citenamefont {Palacci}\ \emph {et~al.}(2013)\citenamefont
  {Palacci}, \citenamefont {Sacanna}, \citenamefont {Steinberg}, \citenamefont
  {Pine},\ and\ \citenamefont {Chaikin}}]{Palacci2013}%
  \BibitemOpen
  \bibfield  {author} {\bibinfo {author} {\bibfnamefont {J.}~\bibnamefont
  {Palacci}}, \bibinfo {author} {\bibfnamefont {S.}~\bibnamefont {Sacanna}},
  \bibinfo {author} {\bibfnamefont {A.~P.}\ \bibnamefont {Steinberg}}, \bibinfo
  {author} {\bibfnamefont {D.~J.}\ \bibnamefont {Pine}},\ and\ \bibinfo
  {author} {\bibfnamefont {P.~M.}\ \bibnamefont {Chaikin}},\ }\bibfield
  {title} {\bibinfo {title} {{Living Crystals of Light-Activated Colloidal
  Surfers}},\ }\href@noop {} {\bibfield  {journal} {\bibinfo  {journal}
  {Science}\ }\textbf {\bibinfo {volume} {339}},\ \bibinfo {pages} {936}
  (\bibinfo {year} {2013})}\BibitemShut {NoStop}%
\bibitem [{\citenamefont {Theurkauff}\ \emph {et~al.}(2012)\citenamefont
  {Theurkauff}, \citenamefont {Cottin-Bizonne}, \citenamefont {Palacci},
  \citenamefont {Ybert},\ and\ \citenamefont {Bocquet}}]{Theurkauff2012}%
  \BibitemOpen
  \bibfield  {author} {\bibinfo {author} {\bibfnamefont {I.}~\bibnamefont
  {Theurkauff}}, \bibinfo {author} {\bibfnamefont {C.}~\bibnamefont
  {Cottin-Bizonne}}, \bibinfo {author} {\bibfnamefont {J.}~\bibnamefont
  {Palacci}}, \bibinfo {author} {\bibfnamefont {C.}~\bibnamefont {Ybert}},\
  and\ \bibinfo {author} {\bibfnamefont {L.}~\bibnamefont {Bocquet}},\
  }\bibfield  {title} {\bibinfo {title} {{Dynamic Clustering in Active
  Colloidal Suspensions with Chemical Signaling}},\ }\href@noop {} {\bibfield
  {journal} {\bibinfo  {journal} {Physical Review Letters}\ ,\ \bibinfo {pages}
  {268303}} (\bibinfo {year} {2012})}\BibitemShut {NoStop}%
\bibitem [{\citenamefont {Ginot}\ \emph {et~al.}(2018)\citenamefont {Ginot},
  \citenamefont {Theurkauff}, \citenamefont {Detcheverry}, \citenamefont
  {Ybert},\ and\ \citenamefont {Cottin-Bizonne}}]{Ginot2018}%
  \BibitemOpen
  \bibfield  {author} {\bibinfo {author} {\bibfnamefont {F.}~\bibnamefont
  {Ginot}}, \bibinfo {author} {\bibfnamefont {I.}~\bibnamefont {Theurkauff}},
  \bibinfo {author} {\bibfnamefont {F.}~\bibnamefont {Detcheverry}}, \bibinfo
  {author} {\bibfnamefont {C.}~\bibnamefont {Ybert}},\ and\ \bibinfo {author}
  {\bibfnamefont {C.}~\bibnamefont {Cottin-Bizonne}},\ }\bibfield  {title}
  {\bibinfo {title} {{Aggregation-fragmentation and individual dynamics of
  active clusters}},\ }\href@noop {} {\bibfield  {journal} {\bibinfo  {journal}
  {Nature Communications}\ }\textbf {\bibinfo {volume} {9}},\ \bibinfo {pages}
  {696} (\bibinfo {year} {2018})}\BibitemShut {NoStop}%
\bibitem [{\citenamefont {Cates}\ and\ \citenamefont
  {Tailleur}(2015)}]{Cates2015}%
  \BibitemOpen
  \bibfield  {author} {\bibinfo {author} {\bibfnamefont {M.~E.}\ \bibnamefont
  {Cates}}\ and\ \bibinfo {author} {\bibfnamefont {J.}~\bibnamefont
  {Tailleur}},\ }\bibfield  {title} {\bibinfo {title} {{Motility-Induced Phase
  Separation}},\ }\href@noop {} {\bibfield  {journal} {\bibinfo  {journal} {The
  Annual Review of Condensed Matter Physics is Annu. Rev. Condens. Matter
  Phys}\ }\textbf {\bibinfo {volume} {6}},\ \bibinfo {pages} {219} (\bibinfo
  {year} {2015})}\BibitemShut {NoStop}%
\bibitem [{\citenamefont {Suma}\ \emph {et~al.}(2014)\citenamefont {Suma},
  \citenamefont {Gonnella}, \citenamefont {Marenduzzo},\ and\ \citenamefont
  {Orlandini}}]{Suma2014}%
  \BibitemOpen
  \bibfield  {author} {\bibinfo {author} {\bibfnamefont {A.}~\bibnamefont
  {Suma}}, \bibinfo {author} {\bibfnamefont {G.}~\bibnamefont {Gonnella}},
  \bibinfo {author} {\bibfnamefont {D.}~\bibnamefont {Marenduzzo}},\ and\
  \bibinfo {author} {\bibfnamefont {E.}~\bibnamefont {Orlandini}},\ }\bibfield
  {title} {\bibinfo {title} {{Motility-induced phase separation in an active
  dumbbell fluid}},\ }\href@noop {} {\bibfield  {journal} {\bibinfo  {journal}
  {EPL (Europhysics Letters)}\ }\textbf {\bibinfo {volume} {108}},\ \bibinfo
  {pages} {56004} (\bibinfo {year} {2014})}\BibitemShut {NoStop}%
\bibitem [{\citenamefont {Petrelli}\ \emph {et~al.}(2018)\citenamefont
  {Petrelli}, \citenamefont {Digregorio}, \citenamefont {Cugliandolo},
  \citenamefont {Gonnella},\ and\ \citenamefont {Suma}}]{Petrelli2018}%
  \BibitemOpen
  \bibfield  {author} {\bibinfo {author} {\bibfnamefont {I.}~\bibnamefont
  {Petrelli}}, \bibinfo {author} {\bibfnamefont {P.}~\bibnamefont
  {Digregorio}}, \bibinfo {author} {\bibfnamefont {L.~F.}\ \bibnamefont
  {Cugliandolo}}, \bibinfo {author} {\bibfnamefont {G.}~\bibnamefont
  {Gonnella}},\ and\ \bibinfo {author} {\bibfnamefont {A.}~\bibnamefont
  {Suma}},\ }\bibfield  {title} {\bibinfo {title} {{Active dumbbells: dynamics
  and morphology in the coexisting region}},\ }\href@noop {} {\bibfield
  {journal} {\bibinfo  {journal} {European Physical Journal E}\ ,\ \bibinfo
  {pages} {128}} (\bibinfo {year} {2018})}\BibitemShut {NoStop}%
\bibitem [{\citenamefont {Klamser}\ \emph {et~al.}(2018)\citenamefont
  {Klamser}, \citenamefont {Kapfer},\ and\ \citenamefont
  {Krauth}}]{Klamser2018}%
  \BibitemOpen
  \bibfield  {author} {\bibinfo {author} {\bibfnamefont {J.~U.}\ \bibnamefont
  {Klamser}}, \bibinfo {author} {\bibfnamefont {S.~C.}\ \bibnamefont
  {Kapfer}},\ and\ \bibinfo {author} {\bibfnamefont {W.}~\bibnamefont
  {Krauth}},\ }\bibfield  {title} {\bibinfo {title} {{Thermodynamic phases in
  two-dimensional active matter}},\ }\href@noop {} {\bibfield  {journal}
  {\bibinfo  {journal} {Nature Communications}\ }\textbf {\bibinfo {volume}
  {9}},\ \bibinfo {pages} {5045} (\bibinfo {year} {2018})}\BibitemShut
  {NoStop}%
\bibitem [{\citenamefont {Levis}\ and\ \citenamefont
  {Berthier}(2014)}]{Levis2014}%
  \BibitemOpen
  \bibfield  {author} {\bibinfo {author} {\bibfnamefont {D.}~\bibnamefont
  {Levis}}\ and\ \bibinfo {author} {\bibfnamefont {L.}~\bibnamefont
  {Berthier}},\ }\bibfield  {title} {\bibinfo {title} {{Clustering and
  heterogeneous dynamics in a kinetic Monte Carlo model of self-propelled hard
  disks}},\ }\href@noop {} {\bibfield  {journal} {\bibinfo  {journal} {Physical
  Review E}\ }\textbf {\bibinfo {volume} {89}},\ \bibinfo {pages} {62301}
  (\bibinfo {year} {2014})}\BibitemShut {NoStop}%
\bibitem [{\citenamefont {Shojaaee}\ \emph {et~al.}(2012)\citenamefont
  {Shojaaee}, \citenamefont {Brendel}, \citenamefont {T\"or\"ok},\ and\
  \citenamefont {Wolf}}]{Zahra2012}%
  \BibitemOpen
  \bibfield  {author} {\bibinfo {author} {\bibfnamefont {Z.}~\bibnamefont
  {Shojaaee}}, \bibinfo {author} {\bibfnamefont {L.}~\bibnamefont {Brendel}},
  \bibinfo {author} {\bibfnamefont {J.}~\bibnamefont {T\"or\"ok}},\ and\
  \bibinfo {author} {\bibfnamefont {D.~E.}\ \bibnamefont {Wolf}},\ }\bibfield
  {title} {\bibinfo {title} {Shear flow of dense granular materials near smooth
  walls. ii. block formation and suppression of slip by rolling friction},\
  }\href@noop {} {\bibfield  {journal} {\bibinfo  {journal} {Phys. Rev. E}\
  }\textbf {\bibinfo {volume} {86}},\ \bibinfo {pages} {011302} (\bibinfo
  {year} {2012})}\BibitemShut {NoStop}%
\bibitem [{\citenamefont {Singh}\ \emph {et~al.}(2020)\citenamefont {Singh},
  \citenamefont {Ness}, \citenamefont {Seto}, \citenamefont {de~Pablo},\ and\
  \citenamefont {Jaeger}}]{Singh2020}%
  \BibitemOpen
  \bibfield  {author} {\bibinfo {author} {\bibfnamefont {A.}~\bibnamefont
  {Singh}}, \bibinfo {author} {\bibfnamefont {C.}~\bibnamefont {Ness}},
  \bibinfo {author} {\bibfnamefont {R.}~\bibnamefont {Seto}}, \bibinfo {author}
  {\bibfnamefont {J.~J.}\ \bibnamefont {de~Pablo}},\ and\ \bibinfo {author}
  {\bibfnamefont {H.~M.}\ \bibnamefont {Jaeger}},\ }\bibfield  {title}
  {\bibinfo {title} {Shear thickening and jamming of dense suspensions: The
  ``roll'' of friction},\ }\href@noop {} {\bibfield  {journal} {\bibinfo
  {journal} {Phys. Rev. Lett.}\ }\textbf {\bibinfo {volume} {124}},\ \bibinfo
  {pages} {248005} (\bibinfo {year} {2020})}\BibitemShut {NoStop}%
\bibitem [{\citenamefont {Nie}\ \emph {et~al.}(2020)\citenamefont {Nie},
  \citenamefont {Chattoraj}, \citenamefont {Piscitelli}, \citenamefont {Doyle},
  \citenamefont {Ni},\ and\ \citenamefont {Ciamarra}}]{Nie2020}%
  \BibitemOpen
  \bibfield  {author} {\bibinfo {author} {\bibfnamefont {P.}~\bibnamefont
  {Nie}}, \bibinfo {author} {\bibfnamefont {J.}~\bibnamefont {Chattoraj}},
  \bibinfo {author} {\bibfnamefont {A.}~\bibnamefont {Piscitelli}}, \bibinfo
  {author} {\bibfnamefont {P.}~\bibnamefont {Doyle}}, \bibinfo {author}
  {\bibfnamefont {R.}~\bibnamefont {Ni}},\ and\ \bibinfo {author}
  {\bibfnamefont {M.~P.}\ \bibnamefont {Ciamarra}},\ }\bibfield  {title}
  {\bibinfo {title} {{Stability phase diagram of active Brownian particles}},\
  }\href@noop {} {\bibfield  {journal} {\bibinfo  {journal} {Physical Review
  Research}\ }\textbf {\bibinfo {volume} {2}},\ \bibinfo {pages} {23010}
  (\bibinfo {year} {2020})}\BibitemShut {NoStop}%
\bibitem [{\citenamefont {Stenhammar}\ \emph {et~al.}(2014)\citenamefont
  {Stenhammar}, \citenamefont {Marenduzzo}, \citenamefont {Allen},\ and\
  \citenamefont {Cates}}]{Stenhammar2014}%
  \BibitemOpen
  \bibfield  {author} {\bibinfo {author} {\bibfnamefont {J.}~\bibnamefont
  {Stenhammar}}, \bibinfo {author} {\bibfnamefont {D.}~\bibnamefont
  {Marenduzzo}}, \bibinfo {author} {\bibfnamefont {R.~J.}\ \bibnamefont
  {Allen}},\ and\ \bibinfo {author} {\bibfnamefont {M.~E.}\ \bibnamefont
  {Cates}},\ }\bibfield  {title} {\bibinfo {title} {{Phase behaviour of active
  Brownian particles: the role of dimensionality}},\ }\href@noop {} {\bibfield
  {journal} {\bibinfo  {journal} {Soft Matter}\ }\textbf {\bibinfo {volume}
  {10}},\ \bibinfo {pages} {1489} (\bibinfo {year} {2014})}\BibitemShut
  {NoStop}%
\bibitem [{\citenamefont {Levis}\ \emph {et~al.}(2017)\citenamefont {Levis},
  \citenamefont {Codina},\ and\ \citenamefont {Pagonabarraga}}]{Levis2017}%
  \BibitemOpen
  \bibfield  {author} {\bibinfo {author} {\bibfnamefont {D.}~\bibnamefont
  {Levis}}, \bibinfo {author} {\bibfnamefont {J.}~\bibnamefont {Codina}},\ and\
  \bibinfo {author} {\bibfnamefont {I.}~\bibnamefont {Pagonabarraga}},\
  }\bibfield  {title} {\bibinfo {title} {{Active Brownian equation of state:
  metastability and phase coexistence}},\ }\href@noop {} {\bibfield  {journal}
  {\bibinfo  {journal} {Soft Matter}\ }\textbf {\bibinfo {volume} {13}},\
  \bibinfo {pages} {8113} (\bibinfo {year} {2017})}\BibitemShut {NoStop}%
\bibitem [{\citenamefont {Digregorio}\ \emph {et~al.}(2018)\citenamefont
  {Digregorio}, \citenamefont {Levis}, \citenamefont {Suma}, \citenamefont
  {Cugliandolo}, \citenamefont {Gonnella},\ and\ \citenamefont
  {Pagonabarraga}}]{Digregorio2018}%
  \BibitemOpen
  \bibfield  {author} {\bibinfo {author} {\bibfnamefont {P.}~\bibnamefont
  {Digregorio}}, \bibinfo {author} {\bibfnamefont {D.}~\bibnamefont {Levis}},
  \bibinfo {author} {\bibfnamefont {A.}~\bibnamefont {Suma}}, \bibinfo {author}
  {\bibfnamefont {L.~F.}\ \bibnamefont {Cugliandolo}}, \bibinfo {author}
  {\bibfnamefont {G.}~\bibnamefont {Gonnella}},\ and\ \bibinfo {author}
  {\bibfnamefont {I.}~\bibnamefont {Pagonabarraga}},\ }\bibfield  {title}
  {\bibinfo {title} {{Full Phase Diagram of Active Brownian Disks: From Melting
  to Motility-Induced Phase Separation}},\ }\href@noop {} {\bibfield  {journal}
  {\bibinfo  {journal} {Physical Review Letters}\ }\textbf {\bibinfo {volume}
  {121}} (\bibinfo {year} {2018})}\BibitemShut {NoStop}%
\bibitem [{\citenamefont {Silbert}(2010)}]{Silbert2010}%
  \BibitemOpen
  \bibfield  {author} {\bibinfo {author} {\bibfnamefont {L.~E.}\ \bibnamefont
  {Silbert}},\ }\bibfield  {title} {\bibinfo {title} {{Jamming of frictional
  spheres and random loose packing}},\ }\href@noop {} {\bibfield  {journal}
  {\bibinfo  {journal} {Soft Matter}\ }\textbf {\bibinfo {volume} {6}},\
  \bibinfo {pages} {2918} (\bibinfo {year} {2010})}\BibitemShut {NoStop}%
\bibitem [{\citenamefont {{Pica Ciamarra}}\ \emph {et~al.}(2011)\citenamefont
  {{Pica Ciamarra}}, \citenamefont {Pastore}, \citenamefont {Nicodemi},\ and\
  \citenamefont {Coniglio}}]{Ciamarra2011}%
  \BibitemOpen
  \bibfield  {author} {\bibinfo {author} {\bibfnamefont {M.}~\bibnamefont
  {{Pica Ciamarra}}}, \bibinfo {author} {\bibfnamefont {R.}~\bibnamefont
  {Pastore}}, \bibinfo {author} {\bibfnamefont {M.}~\bibnamefont {Nicodemi}},\
  and\ \bibinfo {author} {\bibfnamefont {A.}~\bibnamefont {Coniglio}},\
  }\bibfield  {title} {\bibinfo {title} {{Jamming phase diagram for frictional
  particles}},\ }\href@noop {} {\bibfield  {journal} {\bibinfo  {journal}
  {Physical Review E}\ }\textbf {\bibinfo {volume} {84}},\ \bibinfo {pages}
  {041308} (\bibinfo {year} {2011})}\BibitemShut {NoStop}%
\bibitem [{\citenamefont {Liebchen}\ and\ \citenamefont
  {L{\"{o}}wen}(2019)}]{Liebchen2019}%
  \BibitemOpen
  \bibfield  {author} {\bibinfo {author} {\bibfnamefont {B.}~\bibnamefont
  {Liebchen}}\ and\ \bibinfo {author} {\bibfnamefont {H.}~\bibnamefont
  {L{\"{o}}wen}},\ }\bibfield  {title} {\bibinfo {title} {{Which interactions
  dominate in active colloids?}},\ }\href@noop {} {\bibfield  {journal}
  {\bibinfo  {journal} {J. Chem. Phys}\ }\textbf {\bibinfo {volume} {150}},\
  \bibinfo {pages} {61102} (\bibinfo {year} {2019})}\BibitemShut {NoStop}%
\end{thebibliography}
%
\end{document}